\documentstyle[12pt,axodraw,amstex]{article}

\newcommand\vek[1]{\mbox{\rmfamily\bfseries\itshape#1}}

\def\rd{{\rm d}}

\begin{document}

\title{Universal finite-size effects in the two-dimensional
asymmetric Coulomb gas on a sphere}

\author{L. {\v S}amaj$^{1,2}$ \\ \\
Laboratoire de Physique Th\'eorique,
Universit\'e de Paris-Sud, \\ 
B\^atiment 210,
91405 Orsay Cedex, France}

\maketitle

\begin{abstract}
We consider an asymmetric version of a two-dimensional
Coulomb gas, made up of two species of pointlike particles with 
positive $+1$ and negative $-1/Q$ $(Q = 1, 2, \ldots)$ charges;
$Q=1$ corresponds to the symmetric two-component plasma and
the limiting case $Q\to\infty$ is related to the one-component plasma.
The system lives on the surface of a sphere, and
it is studied in both canonical and grand-canonical ensembles.
By combining the method of stereographic projection of the sphere
onto an infinite plane with the technique of a renormalized Mayer
series expansion it is explicitly shown that the finite-size 
expansions of the free energy and of the grand potential 
have the same universal term, independent of model's details.
As a by-product, the collapse temperature and the Kosterlitz-Thouless 
transition point (in the limit of a vanishing hard-core attached to 
particles) are conjectured for any value of $Q$. 
\end{abstract}

\medskip
\noindent LPT Orsay 01-26
\medskip
%PACS numbers: 52.25.Kn, 61.20.Gy, 05.90.+m

\vfill

\noindent $^1$ On leave from the Institute of Physics, Slovak
Academy of Sciences, Bratislava, Slovakia

\makeatletter
\noindent $^2$ E-mail addresses: Ladislav.Samaj@th.u-psud.fr
and fyzimaes@savba.sk
\makeatother

\newpage

\renewcommand{\theequation}{1.\arabic{equation}}
\setcounter{equation}{0}

\section{Introduction}
The universal finite-size properties of two-dimensional (2D)
critical systems with {\em short-range} interactions among
constituents are well understood
within the principle of conformal invariance \cite{Blote}-\cite{Cardy3}.
For a finite 2D system of characteristic size $R$ at critical
point, the dimensionless grand potential $\beta \Omega$ 
[$\beta=1/(kT)$ is the inverse temperature] 
has a large-$R$ expansion of the form
\begin{equation} \label{1.1}
\beta \Omega = A R^2 + B R - {c \chi \over 6}
\ln R + \ldots \ .
\end{equation}
The coefficients $A$ and $B$ of the bulk and surface parts
are non-universal.
The coefficient of the $\ln R$-term is universal, dependent
only on the conformal anomaly number $c$ of the critical theory
and on the Euler number $\chi$ of the manifold (with a smooth
metric and a smooth boundary) to which the system is confined.
In general, $\chi = 2 - 2h - b$, where $h$ is the number of
handles and $b$ the number of boundaries of the manifold
($\chi = 2$ for a sphere, $\chi = 1$ for a disk, $\chi = 0$
for an annulus or a torus).

For lattice models and their continuum limits at critical
point, the quantity (\ref{1.1}) is usually called the 
dimensionless free energy, with the notation $\beta F$,
but from a rigorous point of view those systems are usually treated
in the grand-canonical ensemble (one sums over all spin configurations,
particle occupations, etc.), with the grand potential
as the generating functional.
Here, we will consider a specific class of continuous fluid 
systems, which exhibit at any temperature the universal finite-size
behaviour of type (\ref{1.1}) in both grand-canonical (fixed fugacity, 
the grand potential) and canonical (fixed particle number density, the 
Helmholtz free energy) ensembles.
Although the two ensembles give the same thermodynamic behaviour
in the thermodynamic limit, their finite-size corrections are
expected to differ from one another.
However, as will be documented in this work, at least for the sphere 
geometry of the confining manifold, the universal finite-size correction 
term is the same in both ensembles.

The fluids under consideration are the 2D Coulomb plasmas of charged 
particles, interacting through a {\em long-range} logarithmic potential.
The one-component version (OCP) of equally, say unit, charged point
particles in a uniform neutralizing background is formally
related to the fractional quantum Hall effect \cite{Francesco}.
The model is exactly solvable at the inverse temperature
$\beta = 2$ \cite{Jancovici1} by mapping onto free fermions.
The symmetric two-component version (TCP) of oppositely $\pm 1$
charged pointlike particles undergoes the collapse of positive-negative 
pairs of charges at $\beta = 2$, which corresponds to the exactly
solvable free-fermion point of the equivalent Thirring model 
\cite{Gaudin}-\cite{Cornu2}.
If a small hard core is attached to the particles in order to prevent
the collapse, the 2D TCP (or Coulomb gas) undergoes the famous 
Kosterlitz-Thouless (K-T) transition \cite{Kosterlitz} from a 
high-temperature conducting phase to a low-temperature insulating phase
at around $\beta=4$.

The long-range tail of the Coulomb potential induces screening.
Due to the screening effect, the bulk charge-charge correlations 
are short-ranged, exponential or even Gaussian.
On the other hand, the same screening effect causes that the
induced electrical-field correlations are long-ranged
\cite{Lebowitz}, \cite{Jancovici2}.
As a consequence, the free energy of the 2D OCP and the
grand potential of the symmetric 2D TCP are supposed to
exhibit, at any temperature of the conducting regime,
a universal finite-size correction of type (\ref{1.1});
this correction is opposite to the one occurring in the massless 
Gaussian field theory \cite{Cardy3}) which has $c=1$.
The arguments for a critical-like behaviour were first
given for Coulomb gases with periodic boundary conditions
\cite{Forrester}, then for Coulomb gases confined to a domain
by plain hard walls \cite{Jancovici3}, by ideal-conductor 
boundaries \cite{Jancovici4}, and finally by ideal-dielectric
boundaries \cite{Jancovici5}, \cite{Tellez}.
In all cases, the explicit checks of the predicted finite-size 
universal behaviour were done at the exactly solvable inverse
temperature $\beta=2$ for both OCP and TCP with various geometries of 
confining domains.

The general derivations of the universal finite-size correction term
in refs. \cite{Forrester}-\cite{Tellez} were based on
plausible, but not rigorously justified, heuristic arguments.
Very recently, a direct derivation was made for the specific case
of Coulomb systems living on the surface of a sphere 
[where, since there is no boundary, $B=0$ in (\ref{1.1})].
By combining the method of stereographic projection of the sphere 
onto an infinite plane with linear response theory (TCP, ref.
\cite{Jancovici6}) or with density functional approach
(OCP, ref. \cite{Jancovici7}), the prefactor to the universal
$\ln R$ correction term was related to the second moment of the
short-range part of the density-density direct correlation
function $c_{SR}$ for the planar system.
This second moment, representing a new sum rule for the corresponding
density correlation function, was evaluated explicitly for both OCP
\cite{Kalinay} and TCP \cite{Jancovici8} by using a renormalized 
Mayer expansion \cite{Friedman}, \cite{Deutsch}, namely by observing
a remarkable ``cancelation property'' 
of specific families of renormalized $c_{SR}$-diagrams.
The obtained results confirm quantitatively the prediction
of conformal invariance (\ref{1.1}) for a critical system 
as we had $c=-1$.

In the present paper, we study a more general asymmetric
2D Coulomb gas made up of two species of pointlike particles with
positive and negative charges of arbitrary strengths $q_{\sigma}$ 
$(\sigma = 1,2)$.
Without any loss of generality, we can choose
\begin{equation} \label{1.2}
q_1 = +1 \quad \quad {\rm and} \quad \quad q_2 = -1/Q 
\end{equation}
with $Q = 1, 2, \ldots$ a positive integer.
The model interpolates between the symmetric TCP $(Q=1)$
and the OCP, obtained as the limiting $Q\to\infty$ case
after subtracting the kinetic energy of the 2-species.
For the asymmetric Coulomb gas confined to a sphere,
by combining the method of stereographic projection
\cite{Jancovici6}, \cite{Jancovici7} with the technique
of the renormalized Mayer expansion \cite{Kalinay}-\cite{Deutsch} 
we prove that the prefactor to the universal $\ln R$-term in 
(\ref{1.1}) does not depend on the parameter $Q$.
The same universal term is obtained in both canonical (free energy) and
grand-canonical (grand potential) ensembles, although in the previous 
studies of Coulomb systems the choice of the ensemble seemed to be 
relevant: the exact solution at $\beta=2$ and
finite-size corrections were obtained strictly in the
canonical ensemble for the OCP and in the grand-canonical
ensemble for the TCP.

From a technical point of view, the asymmetry in the strength
of particle charges induces an asymmetry in particle correlations:
two independent two-particle combinations $11 = 22$ and $12 = 21$
in the symmetric TCP are replaced by the three ones
$11$, $22$ and $12 = 21$.
The method of stereographic projection \cite{Jancovici6}, 
\cite{Jancovici7} is shown to imply a new ``missing''
interrelation among the second moments of species pair correlations.
This, together with a generalization of the ``cancelation''
phenomena in the renormalized Mayer expansion, helps us
to extend new sum rules for the plane OCP \cite{Kalinay} and
symmetric TCP \cite{Jancovici8} to the general asymmetric
plasma and thus to find the prefactor to the universal finite-size
correction for the latter model.

In their pioneering work about the asymmetric 2D Coulomb
gas \cite{Hansen1}, the authors predicted a successive
series of temperatures at which partial recombinations (or collapses)
of positive-negative pairs of charges occur, with a consequent
impact on the short-distance behaviour of like-species 
correlations and on the equation of state (discontinuities of 
$\partial p / \partial \beta$, $p$ is the pressure, were suggested).
However, the exact solutions of thermodynamics for the symmetric
$Q=1$ \cite{Samaj1} and the first asymmetric $Q=2$ \cite{Samaj2} cases 
give only one singularity for the pressure at one collapse temperature, 
related to a short-distance instability of the configuration integral 
for a neutral cluster formed of one particle of species 1 and 
$Q$ particles of species 2.
The collapse temperature is explicitly derived for any $Q$
in this work.
If one attach a hard-core $a$ to the particles, the system
remains stable also below the collapse temperature, and can
reach the K-T phase transition.
In the limit of vanishing $a$, the presence of the K-T transition
in the symmetric TCP is indicated by a divergence of the second moment 
of truncated density-density correlations \cite{Jancovici6}.
Having at disposal the generalization of this second moment
sum rule, we suggest 
the position of the  K-T transition temperature for any value of $Q$.

The paper is organized as follows.
The method of stereographic projection in the grand-canonical 
\cite{Jancovici6} and canonical \cite{Jancovici7} ensembles
is briefly generalized to the asymmetric 2D Coulomb gas
in sections 2 and 3, respectively.
Section 4 deals with the renormalized Mayer expansion.
The final results, conjectures and some concluding remarks are given 
in section 5.

\renewcommand{\theequation}{2.\arabic{equation}}
\setcounter{equation}{0}

\section{Grand-canonical ensemble}
On the surface of a sphere of radius $R$, two particles
$i$ of species $\sigma_i$ and $j$ of species $\sigma_j$
interact via the Coulomb interaction 
$ - q_{\sigma_i} q_{\sigma_j} \ln [(2R/L) \sin(\theta_{ij}/2)]$,
where $L$ is a length constant set for simplicity to unity and 
$\theta_{ij}$ is the angular distance of particles \cite{Caillol}.
The particle charges belong to the two-component set 
given by equation (\ref{1.2}).
Under the neutrality condition, the grand partition function reads
\begin{eqnarray} \label{2.1}
\Xi & = & 1 + z_1 (z_2)^Q R^{2(1+Q)} \nonumber \\
& & \times 
\int {\rm d} \Omega_0^{(1)}
{\rm d} \Omega_1^{(2)} \ldots {\rm d}\Omega_Q^{(2)} 
{\prod_{(i<j)=1}^Q \left( 2 R \sin{\theta_{ij}\over 2}
\right)^{\beta/Q^2} \over \prod_{j=1}^Q \left( 2 R
\sin{\theta_{0j}\over 2} \right)^{\beta/Q}} + \ldots \ ,
\end{eqnarray} 
where $z_{\sigma}$ $(\sigma = 1,2)$ are the species fugacities
and ${\rm d} \Omega_i^{(\sigma)}$ is an angle element around
the position of particle $i$ of type $\sigma$.
The term in (\ref{2.1}) containing $N$ particles of type 1
and $QN$ particles of type 2 equals
$[z_1 z_2^Q R^{2(1+Q)(1-{\beta \over 4Q})}]^N$ times
a dimensionless integral dependent only on $\beta$ and $Q$.
Thus, $\ln \Xi$ obeys the homogeneity relations
\begin{equation} \label{2.2}
z_1 {\partial \over \partial z_1} \ln \Xi
= {1\over Q} z_2 {\partial \over \partial z_2} \ln \Xi
= {1\over 2(1+Q)} \left( 1 - {\beta \over 4Q} \right)^{-1}
R {\partial \ln \Xi \over \partial R} \ .
\end{equation}
Let us denote by $N_{\sigma}^{(S)}$ the number of particles
of type $\sigma$ on the sphere of radius $R$; $N_{\sigma}^{(S)} =
4 \pi R^2$ $\times$ the constant density $n_{\sigma}^{(S)}$. 
Since
\begin{equation} \label{2.3}
N_{\sigma}^{(S)} = z_{\sigma} {\partial \over \partial z_{\sigma}}
\ln \Xi \ , \quad \quad \sigma = 1, 2 
\end{equation}
one gets the relations
\begin{subequations} \label{2.4}
\begin{eqnarray}
n_1^{(S)} & = & {1\over 8 \pi (1+Q)} \left( 1 - {\beta \over 4Q}
\right)^{-1} {1\over R} {\partial \over \partial R} \ln \Xi \ ,
\label{2.4a} \\
n_2^{(S)} & = & {Q\over 8 \pi (1+Q)} \left( 1 - {\beta \over 4Q}
\right)^{-1} {1\over R} {\partial \over \partial R} \ln \Xi \ ,
\label{2.4b}
\end{eqnarray}
\end{subequations}
satisfying the electroneutrality condition 
$\sum_{\sigma} q_{\sigma} n_{\sigma}^{(S)} = 0$. 

For fixed fugacities $z_{\sigma}$ and in the large-$R$ limit,
$\ln \Xi$ behaves like
\begin{equation} \label{2.5}
\ln \Xi = \beta p \left( 4\pi R^2 \right) + f(R) + \ldots  \ ,
\end{equation}
where $p$ is the bulk pressure of an infinite (plane) system
and $f(R)$ is the leading finite-size correction.
Substituting (\ref{2.5}) into (\ref{2.4}) one obtains,
in the limit $R\to\infty$,
\begin{subequations} \label{2.6}
\begin{eqnarray}
n_1 & = & {1\over 1+Q} \left( 1 - {\beta \over 4Q}
\right)^{-1} \beta p  \ , \label{2.6a} \\
n_2 & = & {Q\over 1+Q} \left( 1 - {\beta \over 4Q}
\right)^{-1} \beta p  \ . \label{2.6b}
\end{eqnarray}
\end{subequations}
where $n_{\sigma}$ is the number density of species $\sigma$
for the plane system.
The equation of state follows immediately,
\begin{equation} \label{2.7}
\beta p = n \left( 1 - {\beta \over 4Q} \right) \ , 
\end{equation}
where $n = \sum_{\sigma} n_{\sigma}$ is the total number density.
The finite-size corrections to the species number densities
then take the form
\begin{subequations} \label{2.8}
\begin{eqnarray}
n_1^{(S)} - n_1 & = & {1\over 8\pi (1+Q)}
\left( 1 - {\beta \over 4Q} \right)^{-1}
{1\over R} {\partial \over \partial R} f(R) \ ,
\label{2.8a} \\
n_2^{(S)} - n_2 & = & {Q\over 8\pi (1+Q)}
\left( 1 - {\beta \over 4Q} \right)^{-1}
{1\over R} {\partial \over \partial R} f(R) \ .
\label{2.8b}
\end{eqnarray}
\end{subequations}

Under stereographic projection of the sphere from its north pole
onto an infinite plane tangent to its south pole ${\vek 0}$, 
an area element $R^2 {\rm d} \Omega$ on the sphere is mapped 
onto ${\rm d}^2r$ on the plane according to
\begin{equation} \label{2.9}
R^2 {\rm d} \Omega = {{\rm d}^2 r \over 
\left( 1 + {r^2\over 4R^2} \right)^2}  \ ,
\end{equation}
and the angular distance $\theta_{ij}$ between two points
on the sphere is expressible through the distance
$\vert {\vek r}_i - {\vek r}_j \vert$ of their projections
on the plane by
\begin{equation} \label{2.10}
2 R \sin {\theta_{ij} \over 2} = { \vert {\vek r}_i
- {\vek r}_j \vert \over 
\left( 1 + {r_i^2 \over 4R^2} \right)^{1/2}
\left( 1 + {r_j^2 \over 4R^2} \right)^{1/2}} \ .
\end{equation}
After the stereographic transformation, (\ref{2.1}) becomes
the grand partition function of the underlying asymmetric
Coulomb gas formulated on an infinite plane: besides
the two-body interactions $ - q_{\sigma_i} q_{\sigma_j}
\ln \vert {\vek r}_i - {\vek r}_j \vert$, there is a
species-dependent one-body potential acting on each particle,
\begin{subequations} \label{2.11}
\begin{eqnarray}
\beta V_1({\vek r}) & = & 2 \left( 1 - {\beta \over 4} \right)
\ln \left( 1 + {r^2 \over 4 R^2} \right) \underset{R\to\infty}{\simeq}
\left( 1 - {\beta \over 4} \right) {r^2 \over 2 R^2} \ ,
\label{2.11a} \\
\beta V_2({\vek r}) & = & 2 \left( 1 - {\beta \over 4 Q^2} \right)
\ln \left( 1 + {r^2 \over 4 R^2} \right) \underset{R\to\infty}{\simeq}
\left( 1 - {\beta \over 4 Q^2} \right) {r^2 \over 2 R^2} \ , 
\label{2.11b}
\end{eqnarray}
\end{subequations}

The density shift of the planar system at the origin ${\vek 0}$,
due to the one-body potentials $V_{\sigma}({\vek r})$, is
given by linear response theory,
\begin{equation} \label{2.12}
\delta n_{\sigma}({\vek 0}) = - \sum_{\sigma'} \int
{\rm d}^2 r \langle {\hat n}_{\sigma}({\vek 0})
{\hat n}_{\sigma'}({\vek r}) \rangle^{{\rm T}} 
\beta V_{\sigma'}({\vek r}) \ ,
\end{equation}
where ${\hat n}_{\sigma}({\vek r}) = \sum_i \delta_{\sigma,\sigma_i}
\delta({\vek r}-{\vek r}_i)$ is the microscopic number density
of species $\sigma$ at ${\vek r}$ and the truncated statistical
average is taken in the unperturbed planar system.
Let us introduce the notation
\begin{equation} \label{2.13}
I_{\sigma\sigma'} = \int {\rm d}^2 r ~ r^2
\langle {\hat n}_{\sigma}({\vek 0})
{\hat n}_{\sigma'}({\vek r}) \rangle^{{\rm T}}
\end{equation}
for the second moment of the species-density correlation functions,
with the obvious symmetry $I_{\sigma\sigma'}=I_{\sigma'\sigma}$.
Since the density on the sphere equals to its projection on the plane
at the south pole ${\vek 0}$, it holds
\begin{equation} \label{2.14}
\delta n_{\sigma}({\vek 0}) = n_{\sigma}^{(S)} -
n_{\sigma} \ .
\end{equation}
Inserting (\ref{2.11}) into (\ref{2.12}) and considering then 
(\ref{2.14}) in (\ref{2.8}), one finally gets
\begin{subequations} \label{2.15}
\begin{eqnarray}
f(R) & = & - C \ln R  \ , \label{2.15a} \\
C & = & 2\pi (1+Q) \left( 1 - {\beta \over 4 Q} \right) \nonumber \\
& & \times \left[ \left( 1 - {\beta \over 4} \right) \left( I_{11}
+ {1\over Q} I_{21} \right) + \left( 1 - {\beta \over 4 Q^2} \right) 
\left( I_{12} + {1\over Q} I_{22} \right) \right] \ . \label{2.15b}
\end{eqnarray}
\end{subequations}
Simultaneously, as a by-product, 
an important interrelation among $I$'s results:
\begin{equation} \label{2.16}
\left( 1 - {\beta \over 4} \right) \left( I_{11} - {1\over Q}
I_{21} \right) = \left( 1 - {\beta \over 4 Q^2} \right) 
\left( {1\over Q} I_{22} - I_{12} \right)  \ .
\end{equation}

\renewcommand{\theequation}{3.\arabic{equation}}
\setcounter{equation}{0}

\section{Canonical ensemble}
In the canonical format, the stereographic projection maps
the homogeneous particle densities $n_{\sigma}^{(S)}$ on
the sphere onto the inhomogeneous ones on the plane,
\begin{equation} \label{3.1}
n_{\sigma}({\vek r}) = { n_{\sigma}^{(S)} 
\over \left( 1 + {r^2\over 4 R^2} \right)^2 } 
\underset{R\to\infty}{\simeq} n_{\sigma}^{(S)} 
\left( 1 - {r^2\over 2R^2} \right) \ .
\end{equation} 
The local chemical potential of particles $\sigma$ on the
plane is given by
\begin{equation} \label{3.2}
\mu_{\sigma}({\vek r}) = {\delta F[n] \over 
\delta n_{\sigma}({\vek r})} \ ,
\end{equation}
where $F$ is the free energy functional.
According to the elementary fluid theory \cite{Hansen2},
\begin{eqnarray} \label{3.3}
\beta {\delta \mu_{\sigma}({\vek r}) 
\over \delta n_{\sigma'}({\vek r}')} & = & 
{\delta^2 \beta F[n] \over \delta n_{\sigma}({\vek r})
\delta n_{\sigma'}({\vek r}')} \nonumber \\
& = & - c_{\sigma\sigma'}({\vek r},{\vek r}')
+{\delta({\vek r}-{\vek r}') \over n_{\sigma}({\vek r})}
\delta_{\sigma,\sigma'} \ ,
\end{eqnarray}
where $c$ stands for the direct correlation function.
The chemical potentials of the asymmetric Coulomb system on the 
sphere coincide with the ones at the origin of the plane system,
$\mu_{\sigma}({\vek 0})$, for the density profiles (\ref{3.1}).

Let us start from a planar asymmetric Coulomb gas with particle
densities $n_{\sigma}^{(S)}$ and switch on the one-body potentials
$V_{\sigma}({\vek r})$ (\ref{2.11}).
According to (\ref{3.1}), the density change at point ${\vek r}$ 
caused by these one-body potentials is
$\delta n_{\sigma}({\vek r}) \simeq - n_{\sigma}^{(S)} r^2/(2 R^2)$.
Using (\ref{3.3}), the induced shift in the chemical potential at 
the origin is
\begin{eqnarray} \label{3.4}
\beta \delta \mu_{\sigma}({\vek 0}) & = & 
\sum_{\sigma'} \int {\rm d}^2 r \left[ - c_{\sigma\sigma'}({\vek r})
+ {\delta({\vek r}) \over n_{\sigma}({\vek 0})} \delta_{\sigma,\sigma'}
\right] \delta n_{\sigma'}({\vek r}) \nonumber \\
& \simeq & {1\over 2 R^2} \sum_{\sigma'} n_{\sigma'}^{(S)}
\int {\rm d}^2 r ~ c_{\sigma\sigma'}({\vek r}) r^2 \ .
\end{eqnarray}
Consequently,
\begin{equation} \label{3.5}
\beta \sum_{\sigma} N_{\sigma}^{(S)} \delta \mu_{\sigma}({\vek 0})
\simeq 2 \pi \sum_{\sigma,\sigma'} n_{\sigma}^{(S)} n_{\sigma'}^{(S)}
\int {\rm d}^2 r ~ c_{\sigma\sigma'}({\vek r}) r^2 \ .
\end{equation}

We will now omit the superscript $(S)$ in the notation and
denote $N = \sum_{\sigma} N_{\sigma}$.
The free energy $F$ has the large-$N$ behaviour
\begin{equation} \label{3.6}
\beta F = A N + g(N) + {\rm const} + \ldots \ .
\end{equation}
Since $\mu_{\sigma} = \partial F / \partial N_{\sigma}$,
\begin{equation} \label{3.7}
\beta \delta\mu_{\sigma} \simeq {\partial g(N) \over \partial N} \ .
\end{equation}
Taking into account that 
$\delta\mu_{\sigma} = \delta\mu_{\sigma}({\vek 0})$,
eq. (\ref{3.5}) then implies
\begin{subequations} \label{3.8}
\begin{eqnarray}
g(N) & = & D \ln N \ , \label{3.8a} \\
D & = & 2 \pi \sum_{\sigma,\sigma'} n_{\sigma} n_{\sigma'}
\int {\rm d}^2 r ~ c_{\sigma\sigma'}({\vek r}) r^2 \ ,
\end{eqnarray}
\end{subequations}
where $n_{\sigma}$ are the particle densities of the asymmetric
Coulomb gas formulated on an infinite plane.

\renewcommand{\theequation}{4.\arabic{equation}}
\setcounter{equation}{0}

\section{Renormalized Mayer expansion}
We first review briefly the renormalized Mayer expansion 
in density (for details, see refs. \cite{Kalinay} - \cite{Deutsch}) 
for a general fluid composed of distinct species of particles
$\{ \sigma \}$ with the corresponding densities 
$\{ n({\vek r},\sigma) \}$.
The particles $i$ and $j$ interact through the pair potential
$v({\vek r}_i,\sigma_i\vert {\vek r}_j,\sigma_j)$;
vector position ${\vek r}_i$ will be sometimes represented
simply by $i$.
The renormalization technique is based on an expansion 
of each Mayer function in the inverse temperature $\beta$, and
on a consequent series resummation of two-coordinated field
circles; by coordination of a circle we mean the number 
of bonds meeting at this circle.
The renormalized $K$-bonds are given by
$$
{\begin{picture}(32,20)(0,7)
    \Photon(0,10)(32,10){1}{7}
    \BCirc(0,10){2.5} \BCirc(32,10){2.5}
    \Text(0,0)[]{$1,\sigma_1$} \Text(32,0)[]{$2,\sigma_2$}
    \Text(17,23)[]{$K$}
\end{picture}}
\ \ \ = \ \ \
{\begin{picture}(32,20)(0,7)
    \DashLine(0,10)(32,10){5}
    \BCirc(0,10){2.5} \BCirc(32,10){2.5}
    \Text(0,0)[]{$1,\sigma_1$} \Text(32,0)[]{$2,\sigma_2$}
    \Text(17,23)[]{$-\beta v$} 
\end{picture}}\ \ +\ \
{\begin{picture}(64,20)(0,7)
    \DashLine(0,10)(32,10){5}
    \DashLine(32,10)(64,10){5}
    \BCirc(0,10){2.5} \BCirc(64,10){2.5}
    \Vertex(32,10){2.2}
    \Text(0,0)[]{$1,\sigma_1$} \Text(64,0)[]{$2,\sigma_2$}
\end{picture}}\ \ +\ \
{\begin{picture}(96,20)(0,7)
    \DashLine(0,10)(32,10){5}
    \DashLine(32,10)(64,10){5}
    \DashLine(64,10)(96,10){5}
    \BCirc(0,10){2.5} \BCirc(96,10){2.5}
    \Vertex(32,10){2.2} \Vertex(64,10){2.2}
    \Text(0,0)[]{$1,\sigma_1$} \Text(96,0)[]{$2,\sigma_2$}
\end{picture}}\ \ + \ldots$$
or, algebraically,
\begin{eqnarray} \label{4.1}
K(1,\sigma_1\vert 2,\sigma_2) & = & 
[-\beta v(1,\sigma_1 \vert 2, \sigma_2)] \nonumber \\
& & + \sum_{\sigma_3} \int [-\beta v(1,\sigma_1 \vert 3,\sigma_3)]~
n(3,\sigma_3)~ K(3,\sigma_3 \vert 2,\sigma_2)~ \rd 3 \ .
\end{eqnarray}
It is straightforward to verify by variation of (\ref{4.1})
that it holds
\begin{equation} \label{4.2}
{\delta K(1,\sigma_1\vert 2,\sigma_2) \over \delta n(3,\sigma_3)}
= K(1,\sigma_1\vert 3,\sigma_3) K(3,\sigma_3\vert 2,\sigma_2) \ .
\end{equation}
The procedure of bond-renormalization transforms the ordinary
Mayer representation of the excess Helmholtz free energy, 
defined by $-\beta F^{{\rm ex}} = - \beta F + \sum_{\sigma} 
\int {\rm d}1~ [ n(1,\sigma) \ln n(1,\sigma) - n(1,\sigma)]$, into
\begin{subequations} \label{4.3}
\begin{equation} \label{4.3a}
- \beta F^{{\rm ex}} = \ \  
\begin{picture}(50,20)(0,7)
    \DashLine(0,10)(40,10){5}
    \Vertex(0,10){2.2} \Vertex(40,10){2.2}
\end{picture}
 + \ \ D^{(0)}[n]  +  \sum_{s=1}^{\infty} D^{(s)}[n] \ ,
\end{equation}
where
\begin{equation} \label{4.3b}
D^{(0)} = \ \
\begin{picture}(40,20)(0,7)
    \DashCArc(20,-10)(28,45,135){5}
    \DashCArc(20,30)(28,225,315){5}
    \Vertex(0,10){2} \Vertex(40,10){2}
\end{picture} \ \ +\ \ 
\begin{picture}(40,20)(0,19)
    \DashLine(0,10)(40,10){5}
    \DashLine(0,10)(20,37){5}
    \DashLine(20,37)(40,10){5}
    \Vertex(0,10){2} \Vertex(40,10){2} \Vertex(20,37){2}
\end{picture} \ \ +\ \ 
\begin{picture}(30,20)(0,10)
    \DashLine(0,0)(30,0){5}
    \DashLine(0,0)(0,30){5}
    \DashLine(0,30)(30,30){5}
    \DashLine(30,0)(30,30){5}
    \Vertex(0,0){2} \Vertex(30,0){2} \Vertex(0,30){2} \Vertex(30,30){2}
\end{picture} \ \ +\ \ \ldots 
\end{equation}
\vskip 0.25cm
\noindent is the sum of all unrenormalized ring diagrams and
\begin{equation} \label{4.3c}
\begin{picture}(60,40)(0,7)
    \PhotonArc(20,6)(20,15,165){1}{11}
    \PhotonArc(20,14)(20,195,345){1}{11}
    \Photon(0,10)(40,10){1}{8.5}
    \Vertex(0,10){2} \Vertex(40,10){2}
    \Text(20,-25)[]{$D^{(1)}$}
\end{picture}
\begin{picture}(60,40)(0,7)
    \PhotonArc(20,-10)(28,45,135){1}{9}
    \PhotonArc(20,30)(28,225,315){1}{9}
    \PhotonArc(20,6)(20,15,165){1}{11}
    \PhotonArc(20,14)(20,195,345){1}{11}
    \Vertex(0,10){2} \Vertex(40,10){2}
    \Text(20,-25)[]{$D^{(2)}$}
\end{picture}
\begin{picture}(60,40)(0,16)
    \PhotonArc(32,6)(32,115,170){1}{7.5}
    \PhotonArc(-12,40)(32,295,355){1}{7.5}
    \PhotonArc(9,5)(32,10,70){1}{7.5}
    \PhotonArc(52,40)(32,185,250){1}{7.5}
    \Photon(0,10)(40,10){1}{8}
    \Vertex(0,10){2} \Vertex(40,10){2} \Vertex(20,35){2}
    \Text(20,-15)[]{$D^{(3)}$}
\end{picture}
\SetScale{0.9}
\begin{picture}(60,40)(0,13)
    \Photon(0,0)(0,40){1}{7}
    \Photon(40,0)(40,40){1}{7}
    \Vertex(0,0){2} \Vertex(40,0){2} \Vertex(0,40){2} \Vertex(40,40){2}
    \PhotonArc(20,-20)(28,45,135){1}{8}
    \PhotonArc(20,20)(28,225,315){1}{8}
    \PhotonArc(20,20)(28,45,135){1}{8}
    \PhotonArc(20,60)(28,225,315){1}{8}
    \Text(20,-19)[]{$D^{(4)}$}
\end{picture} 
\begin{picture}(60,40)(0,12)
    \Photon(0,0)(40,0){1}{7}
    \Photon(0,0)(0,40){1}{7}
    \Photon(0,40)(40,40){1}{7}
    \Photon(40,0)(40,40){1}{7}
    \Photon(0,0)(40,40){1}{8}
    \Photon(0,40)(40,0){1}{8}
    \Vertex(0,0){2} \Vertex(40,0){2} \Vertex(0,40){2} \Vertex(40,40){2}
    \Text(20,-19)[]{$D^{(5)}$}
\end{picture}
\SetScale{1}  , \  {\rm etc.}
\end{equation}
\vskip 0.80cm
\end{subequations}
\noindent represent the set of all remaining
completely renormalized graphs.
The free energy is the generating functional for the direct
correlation function $c$ in the sense that
\begin{equation} \label{4.4}
c(1,\sigma_1\vert 2,\sigma_2) = {\delta^2 (-\beta F^{{\rm ex}}) \over
\delta n(1,\sigma_1) \delta n(2,\sigma_2)} \ .
\end{equation}
With regard to (\ref{4.3}),
\begin{subequations} \label{4.5}
\begin{equation} \label{4.5a}
c(1,\sigma_1\vert 2,\sigma_2) = 
-\beta v(1,\sigma_1\vert 2,\sigma_2)
+ {1\over 2!} K^2(1,\sigma_1\vert 2,\sigma_2)
+ \sum_{s=1}^{\infty} c^{(s)}(1,\sigma_1\vert 2,\sigma_2) \ ,
\end{equation}
where $c^{(s)}(1,\sigma_1\vert 2,\sigma_2)$ represents
the whole family of $(1,\sigma_1)(2,\sigma_2)$-rooted diagrams
generated from a given $D^{(s)}$,
\begin{equation} \label{4.5b}
c^{(s)}(1,\sigma_1\vert 2,\sigma_2) = {\delta^2 D^{(s)} \over 
\delta n(1,\sigma_1) \delta n(2,\sigma_2)} \ .
\end{equation}
\end{subequations}
To get explicitly $c^{(s)}$, one has
to take into account that the functional derivative of 
$D^{(s)}$ with respect to the density field generates
root circles not only at obvious field-circle positions,
but also on $K$ bonds according to formula (\ref{4.2}),
causing their right $K$-$K$ division.
The pair distribution function $h$, defined by
\begin{equation} \label{4.6}
n(1,\sigma_1) n(2,\sigma_2) h(1,\sigma_1\vert 2,\sigma_2)
= \langle {\hat n}_{\sigma_1}(1) {\hat n}_{\sigma_2}(2)\rangle^{{\rm T}} 
- \delta_{\sigma_1,\sigma_2} \delta(1-2) n(1,\sigma_1) \ ,
\end{equation}
is related to $c$ via the Ornstein-Zernike (OZ) equation
\begin{eqnarray} \label{4.7}
h(1,\sigma_1\vert 2,\sigma_2) & = & c(1,\sigma_1\vert 2,\sigma_2)
\nonumber \\
& & + \sum_{\sigma_3} \int {\rm d} 3 ~ c(1,\sigma_1\vert 3,\sigma_3)
n(3,\sigma_3) h(3,\sigma_3\vert 2,\sigma_2) \ .
\end{eqnarray}

Let us return to the 2D asymmetric Coulomb gas with two kinds of 
particles $\sigma = 1,2$ of charges (\ref{1.2}), interacting via 
the logarithmic interaction
\begin{subequations} \label{4.8}
\begin{eqnarray} 
v(i,\sigma_i\vert j,\sigma_j) & = & q_{\sigma_i} q_{\sigma_j}
v(i,j) \ , \label{4.8a} \\
v(i,j) & = & - \ln \vert i - j \vert \ . \label{4.8b}
\end{eqnarray} 
\end{subequations}
We consider the infinite-volume limit, characterized by
homogeneous densities $n(1,\sigma) = n_{\sigma}$ 
constrained by the neutrality condition 
$\sum_{\sigma} q_{\sigma} n_{\sigma} = 0$,
so that 
\begin{equation} \label{4.9}
n_1 = n {1 \over 1+Q} \ , \quad \quad n_2 =  n {Q \over 1+Q} \ ,
\end{equation} 
with $n$ being the total particle density.
Two-body functions are both isotropic and translationally invariant, 
$c(1,\sigma_1\vert 2,\sigma_2) = c_{\sigma_1 \sigma_2}(\vert 1-2\vert)$,
$h(1,\sigma_1\vert 2,\sigma_2) = h_{\sigma_1 \sigma_2}(\vert 1-2\vert)$.
From eq. (\ref{4.1}) it follows that the renormalized bonds 
exhibit the same charge-dependence as the 
interaction under consideration (\ref{4.8}),
\begin{subequations} \label{4.10}
\begin{eqnarray} 
K(1,\sigma_1\vert 2,\sigma_2) & = & q_{\sigma_1} q_{\sigma_2} K(1,2)
\ , \label{4.10a} \\
K(1,2) & = & - \beta K_0(\kappa \vert 1-2 \vert) \ . \label{4.10b}
\end{eqnarray}
\end{subequations}
Here, $K_0$ is the modified Bessel function of second kind
and $\kappa$ is the inverse Debye length defined by
\begin{equation} \label{4.11}
\kappa^2 = 2 \pi \beta \sum_{\sigma} q_{\sigma}^2 n_{\sigma} \ .
\end{equation}
The direct correlation function $c$, given by (\ref{4.5a}), reads
\begin{equation} \label{4.12}
c_{\sigma_1\sigma_2}(\vert 1-2\vert)  =  q_{\sigma_1} q_{\sigma_2}
\beta \ln \vert 1-2 \vert + q_{\sigma_1}^2 q_{\sigma_2}^2 {1\over 2!}
\left[ \beta K_0(\kappa\vert 1-2\vert) \right]^2 + \sum_{s=1}^{\infty} 
c_{\sigma_1\sigma_2}^{(s)}(\vert 1-2 \vert)
\end{equation}

For renormalized bonds of type (\ref{4.10}),
the $c^{(s)}$-families of diagrams exhibit, as separate units,
remarkable ``cancelation properties'', regardless of 
the topology of the generating graph $D^{(s)}$.
Let the given completely renormalized diagram $D^{(s)}$
$(s=2,1,\ldots)$ be composed of $U$ skeleton vertices
$i = 1, \ldots, U$ of bond coordinations $\{ \nu_i \}$
and $V = \sum_{i=1}^U \nu_i/2$ bonds $a = 1, \ldots, V$.
Omitting topological factors, $D^{(s)}$ can be formally
expressed as
\begin{equation} \label{4.13}
D^{(s)}[n] = \int \prod_{i=1}^U \left[ {\rm d}i \sum_{\sigma_i}
n(i,\sigma_i) \right] \prod_{a=1}^V K(a_1,\sigma_{a_1}\vert
a_2,\sigma_{a_2}) \ ,
\end{equation}
where $a_1, a_2 \in \{ 1, \ldots, U \}, a_1<a_2$, denote
the ordered pair of vertices joint by the $a$-bond.
For the family $c^{(s)}$, generated from $D^{(s)}$
according to (\ref{4.5b}), one gets
\begin{eqnarray} \label{4.14}
& & \sum_{\sigma,\sigma'} n_{\sigma} n_{\sigma'}
c_{\sigma\sigma'}^{(s)}(\vert {\vek r}-{\vek r}'\vert)
= \prod_{i=1}^U \left( \sum_{\sigma_i} q_{\sigma_i}^{\nu_i}
n_{\sigma_i} \right) \nonumber \\
(a) \hskip 0.8truecm &\times & \Big[ \int \prod_i {\rm d}i
\sum_{i,j\atop (i\ne j)} \delta({\vek r}-i) \delta({\vek r}'-j)
\prod_a K_a  \nonumber \\
(b) \hskip 0.8truecm 
& & + {\tilde n} \int \prod_i {\rm d}i \sum_i \delta({\vek r}-i)
\sum_a K(a_1,{\vek r}') K({\vek r}',a_2) \prod_{b\ne a} K_b
\nonumber \\
(c) \hskip 0.8truecm 
& & + {\tilde n} \int \prod_i {\rm d}i \sum_i \delta({\vek r}'-i)
\sum_a K(a_1,{\vek r}) K({\vek r},a_2) \prod_{b\ne a} K_b
\nonumber \\
(d) \hskip 0.8truecm 
& & + ({\tilde n})^2 \int \prod_i {\rm d}i \sum_a K(a_1,{\vek r}')
K({\vek r}',{\vek r}) K({\vek r},a_2) \prod_{b\ne a} K_b 
\nonumber \\
(e) \hskip 0.8truecm 
& & + ({\tilde n})^2 \int \prod_i {\rm d}i \sum_a K(a_1,{\vek r})
K({\vek r},{\vek r}') K({\vek r}',a_2) \prod_{b\ne a} K_b 
\nonumber \\
(f) \hskip 0.8truecm 
& & + ({\tilde n})^2 \int \prod_i {\rm d}i \sum_{a,b\atop (a\ne b)} 
K(a_1,{\vek r}) K({\vek r},a_2) K(b_1,{\vek r}') K({\vek r}',b_2)
\prod_{c\ne a,b} K_c \Big] \ , 
\nonumber \\
\end{eqnarray}
where we have applied the functional relation (\ref{4.2}).
Here, $K_a \equiv K(a_1,a_2)$ defined by (\ref{4.10a}) and
${\tilde n} = \sum_{\sigma}q_{\sigma}^2 n_{\sigma}= n/Q$.
The $(a)$ term in (\ref{4.14}) corresponds to the creation
of root points at two skeleton vertices, the next $(b,c)$ terms
to one root circle generated at the skeleton and the other one
at a bond, the $(d,e)$ terms to two root points at the same
bond and the last $(f)$ term represents root points generated
at two different renormalized bonds.
Comparing (\ref{4.14}) with formula (31) of ref. \cite{Kalinay}
and assuming that $\kappa^2 = 2\pi \beta {\tilde n}$,
it becomes clear that the rhs of (\ref{4.14}) is proportional
to $c^{(s)}(\vert {\vek r}-{\vek r}'\vert)$ generated from
$D^{(s)}$ for the case of the 2D OCP with mobile-particle density
${\tilde n}$.
As was proven in ref. \cite{Kalinay}, the zeroth and second
moments of such a $c^{(s)}$ vanish, regardless of the topology
of the generating diagram $D^{(s)}$.
Consequently, in our model,
\begin{subequations} \label{4.15}
\begin{eqnarray}
\sum_{\sigma,\sigma'} n_{\sigma} n_{\sigma'}
\int {\rm d}^2 r ~ c_{\sigma\sigma'}^{(s)}({\vek r}) & = & 0
\ , \label{4.15a} \\
\sum_{\sigma,\sigma'} n_{\sigma} n_{\sigma'}
\int {\rm d}^2 r ~ r^2 c_{\sigma\sigma'}^{(s)}({\vek r}) & = & 0
\ , \label{4.15b}
\end{eqnarray}
\end{subequations}
for every $s = 1, 2, \ldots$.

The 2D Fourier transform is defined by
\begin{subequations} \label{4.16}
\begin{eqnarray}
f({\vek r}) & = & {1\over 2\pi} \int {\rm d}^2 k ~
{\bar f}({\vek k}) \exp({\rm i} {\vek k} \cdot {\vek r}) \ ,
\label{4.16a} \\
{\bar f}({\vek k}) & = & {1\over 2\pi} \int {\rm d}^2 r ~
f({\vek r}) \exp(- {\rm i} {\vek k} \cdot {\vek r}) \ .
\label{4.16b}
\end{eqnarray}
\end{subequations}
For a function with circular symmetry, whose Fourier component
is analytic in $k$ around $k=0$, one has
\begin{subequations}
\begin{eqnarray} \label{4.17}
{\bar f}(k) & = & \sum_{j=0}^{\infty} {\bar f}^{(2j)} k^{2j} \ , 
\label{4.17a} \\
{\bar f}^{(2j)} & = &  {(-1)^j \over 4^j (j!)^2} {1\over 2\pi}
\int {\rm d}^2 r ~ r^{2j} f(r) \ . \label{4.17b}
\end{eqnarray}
\end{subequations}
The Fourier component of $K_0^2(r)$ is given by
\begin{eqnarray} \label{4.18}
\int {{\rm d}^2 r \over 2\pi} K_0^2(r) 
\exp \left( - i {\vek k}\cdot {\vek r} \right)
& = & \int_0^{\infty} {\rm d} r ~ r J_0(k r) K_0^2(r) \nonumber \\
& = & {\ln \left( k/2 + \sqrt{1+(k/2)^2} \right) \over 
k \sqrt{1 + (k/2)^2} } \ ,
\end{eqnarray}
with $J_0$ being the ordinary Bessel function.
The Fourier transform of the direct correlation function $c$,
eq. (4.12), thus reads
\begin{equation} \label{4.19}
{\bar c}_{\sigma\sigma'}(k) = - q_{\sigma} q_{\sigma'} {\beta \over k^2}
+ {\beta^2 \over 2 \kappa} q_{\sigma}^2 q_{\sigma'}^2~ { \ln \left(
k/(2\kappa) + \sqrt{ 1 + [k/(2\kappa)]^2 } \right) \over k ~
\sqrt{1+[k/(2\kappa)]^2}} + \sum_{s=1}^{\infty} 
{\bar c}_{\sigma\sigma'}^{(s)}(k)
\end{equation}
where ${\bar c}_{\sigma\sigma'}^{(s)}(k)$ are analytic functions of $k$.
Regarding the neutrality condition, relations (\ref{4.15})
can be reflected via
\begin{equation} \label{4.20}
\sum_{\sigma,\sigma'} n_{\sigma} n_{\sigma'} 
{\bar c}_{\sigma\sigma'}(k) = {\kappa^2 \over 16 \pi^2}
- {k^2 \over 96 \pi^2} + O(k^4)
\end{equation} 
Note that the coefficient to the $k^2$-term is universal, i.e.,
independent of $\beta$, $n_1$ and $n_2$.
In ${\vek k}$-space, the OZ equation (4.7) takes the form
\begin{equation} \label{4.21}
{\bar h}_{\sigma\sigma'}(k) = {\bar c}_{\sigma\sigma'}(k)
+ 2 \pi \sum_{\sigma''} {\bar c}_{\sigma\sigma''}(k) n_{\sigma''}
{\bar h}_{\sigma''\sigma'}(k)
\end{equation} 

The insertion of the leading term 
${\bar c}_{\sigma\sigma'}(k) \simeq - q_{\sigma} 
q_{\sigma'} \beta / k^2$ into (\ref{4.21}) leads to the
zeroth-moment (electroneutrality)
\begin{equation} \label{4.22}
\sum_{\sigma'} q_{\sigma'} n_{\sigma'} 
{\bar h}_{\sigma'\sigma}^{(0)} = - {q_{\sigma} \over 2\pi} \ ,
\quad \quad \sigma = 1, 2
\end{equation}
and the second-moment
\begin{equation} \label{4.23}
\sum_{\sigma,\sigma'} q_{\sigma} n_{\sigma} q_{\sigma'} n_{\sigma'}
{\bar h}_{\sigma\sigma'}^{(2)} = {1\over (2\pi)^2 \beta}
\end{equation}
Stillinger-Lovett sum rules \cite{Stillinger1}, \cite{Stillinger2}
for the truncated {\em charge-charge} correlation function
$\langle {\hat\rho}({\vek 0}) {\hat\rho}({\vek r}) \rangle^{{\rm T}}$,
where ${\hat\rho}({\vek r}) = \sum_{\sigma} q_{\sigma} 
{\hat n}_{\sigma}({\vek r})$ is the total microscopic charge
density at ${\vek r}$ [note the definition of $h$ in (\ref{4.6})].

To make use of (\ref{4.20}), we introduce an auxilliary function 
\begin{equation} \label{4.24}
{\bar f}_{\sigma}(k) = \sum_{\sigma'} n_{\sigma'}
{\bar c}_{\sigma'\sigma}(k)
\end{equation}
satisfying the relations
\begin{subequations} \label{4.25}
\begin{eqnarray}
n_1 {\bar f}_1^{(0)} + n_2 {\bar f}_2^{(0)} & = & 
{\beta n \over 8 \pi Q} \ ,
\label{4.25a} \\
n_1 {\bar f}_1^{(2)} + n_2 {\bar f}_2^{(2)} & = & 
- {1\over 96 \pi^2} \ .
\label{4.25b}
\end{eqnarray}
\end{subequations}

At the lowest $k^0$ order, the OZ equation (\ref{4.21}) gives
\begin{equation} \label{4.26}
\sum_{\sigma'} n_{\sigma'} {\bar h}_{\sigma'\sigma}^{(0)}
= {\bar f}_{\sigma}^{(0)} + 2\pi \sum_{\sigma'} 
{\bar f}_{\sigma'}^{(0)} n_{\sigma'} {\bar h}_{\sigma'\sigma}^{(0)} \ . 
\end{equation}
Combining (\ref{4.26}) with (\ref{4.22}), ${\bar f}_1^{(0)}$ and
${\bar f}_2^{(0)}$ occur only in the combination (\ref{4.25a}),
so they can be eliminated.
Consequently, one gets explicitly
\begin{subequations} \label{4.27}
\begin{eqnarray}
{\bar h}_{1 1}^{(0)} & = & {1\over 2\pi n [ 1 - \beta/(4Q) ]}
- {1+Q \over 2\pi n } \ , \label{4.27a} \\
{\bar h}_{2 2}^{(0)} & = & {1\over 2\pi n [ 1 - \beta/(4Q) ]}
- {1+Q \over 2\pi n Q } \ , \label{4.27b} \\
{\bar h}_{1 2}^{(0)} & = & {1\over 2\pi n [ 1 - \beta/(4Q) ]} \ . 
\label{4.27c}
\end{eqnarray}
\end{subequations}
Without going into details, these formulae imply the
compressibility sum rule \cite{Hansen1}, \cite{Vieillefosse}, i.e.,
the zeroth moment of the truncated {\em density-density} correlation 
function $\langle {\hat n}({\vek 0}) {\hat n}({\vek r}) \rangle^{{\rm T}}$,
where ${\hat n}({\vek r}) = \sum_{\sigma} {\hat n}_{\sigma}({\vek r})$ 
is the total microscopic number density at ${\vek r}$.

At the next $k^2$ order, the OZ equation (\ref{4.21}) gives
\begin{equation} \label{4.28}
\sum_{\sigma'} n_{\sigma'} {\bar h}_{\sigma'\sigma}^{(2)}
=  {\bar f}_{\sigma}^{(2)} + 2\pi \sum_{\sigma'} {\bar f}_{\sigma'}^{(2)}
n_{\sigma'} {\bar h}_{\sigma'\sigma}^{(0)} 
+ 2\pi \sum_{\sigma'} {\bar f}_{\sigma'}^{(0)} n_{\sigma'}
{\bar h}_{\sigma'\sigma}^{(2)}
\end{equation}
Inserting (\ref{4.27}) into (\ref{4.28}), ${\bar f}_1^{(2)}$ and
${\bar f}_2^{(2)}$ occur only in the combination (\ref{4.25b}),
so they can be eliminated.
One obtains
\begin{subequations} \label{4.29}
\begin{eqnarray}
{\bar h}_{1 1}^{(2)} + Q {\bar h}_{2 1}^{(2)} & = & 
- {1\over 96 \pi^2 n^2} {1+Q \over 1-\beta/(4Q)}
+ 2 \pi \left( {\bar f}_1^{(0)} {\bar h}_{1 1}^{(2)}
+ Q {\bar f}_2^{(0)} {\bar h}_{2 1}^{(2)} \right) \ , \label{4.28a} \\
{\bar h}_{1 2}^{(2)} + Q {\bar h}_{2 2}^{(2)} & = & 
- {1\over 96 \pi^2 n^2} {1+Q \over 1-\beta/(4Q)}
+ 2 \pi \left( {\bar f}_1^{(0)} {\bar h}_{1 2}^{(2)}
+ Q {\bar f}_2^{(0)} {\bar h}_{2 2}^{(2)} \right) \ . \label{4.28b}
\end{eqnarray}
\end{subequations}
From eqs. (\ref{4.6}) and (\ref{4.17}) it follows that the
second moments $I_{\sigma\sigma'}$, defined by (\ref{2.13}),
are related to ${\bar h}_{\sigma\sigma'}^{(2)}$ in the following
way
\begin{equation} \label{4.30}
I_{\sigma\sigma'} = - 8 \pi n_{\sigma} n_{\sigma'}
{\bar h}_{\sigma\sigma'}^{(2)} \ .
\end{equation}
From the complete set of equations (\ref{4.29}), (\ref{4.25a}),
(\ref{4.23}) and the one (\ref{2.16}) obtained by the stereographic 
projection, one finally finds that ${\bar f}_1^{(0)} = \beta / (8 \pi)$,
${\bar f}_2^{(0)} = \beta / (8 \pi Q^2)$ and
\begin{subequations} \label{4.31}
\begin{eqnarray}
I_{1 1} & = & - {2 (3\beta - 8 Q^2) (\beta - 6 Q^2) \over
3 \beta \pi (\beta - 4 Q)^2 (1+Q)^2} \ , \label{4.31a} \\
I_{2 2} & = & - {2 Q^4 (3\beta - 8) (\beta - 6) \over
3 \beta \pi (\beta - 4 Q)^2 (1+Q)^2} \ , \label{4.31b} \\
I_{1 2} & = & {2 Q^2 \left[ 3\beta^2 - 2 \beta (6 - Q + 6 Q^2) 
+ 48 Q^2 \right] \over
3 \beta \pi (\beta - 4 Q)^2 (1+Q)^2} \ . \label{4.31c}
\end{eqnarray}
\end{subequations}
With regard to the definition of $I$'s (\ref{2.13}),
the second moment of the number density correlation function is
\begin{equation} \label{4.32}
\int {\rm d}^2 r ~ r^2 \langle {\hat n}({\vek 0}) 
{\hat n}({\vek r}) \rangle^{{\rm T}}
= {1-3\beta (Q-1)^2/(2 Q^2) \over 12 \pi [1-\beta/(4Q)]^2}
\end{equation}
This is the generalization of the new sum rule derived for the
symmetric $Q=1$ TCP in refs. \cite{Jancovici6} and \cite{Jancovici7}.

\renewcommand{\theequation}{5.\arabic{equation}}
\setcounter{equation}{0}

\section{Final results and conclusion}
Inserting (\ref{4.31}) into (\ref{2.15b}), one gets $C=1/3$.
With regard to (\ref{2.5}) and (\ref{2.15a}), the dimensionless
grand potential, $\beta \Omega = - \ln \Xi$, has the large-$R$
expansion
\begin{equation} \label{5.1}
\beta \Omega = - \beta p \left( 4\pi R^2 \right) +
{1\over 3} \ln R + {\rm const} + \ldots
\end{equation}  
This expansion has the universal $\ln R$ term like in (\ref{1.1}).
Taking the sphere value of $\chi = 2$, its prefactor corresponds
to the conformal anomaly number $c = -1$ as was expected.

With regard to (\ref{4.17}), it follows from (\ref{4.20}) that
\begin{equation} \label{5.2}
\sum_{\sigma,\sigma'} n_{\sigma} n_{\sigma'} \int {\rm d}^2 r
~ c_{\sigma\sigma'}(r) r^2 = {1\over 12 \pi} .
\end{equation}
Consequently, $D=1/6$ in (\ref{3.8}), and (\ref{3.6}) takes
the form
\begin{equation} \label{5.3}
\beta F = A N + {1\over 6} \ln N + {\rm const} + \ldots
\end{equation}
Since $N=4\pi R^2 n$, the dimensionless free energy $\beta F$
has the same universal correction term as $\beta \Omega$.
Relations (\ref{5.1}) and (\ref{5.3}), valid for any asymmetric 
2D Coulomb gas, are the main results of the present work.

As was promised in the Introduction, we now derive an expression
for the collapse temperature as a function of $Q$.
On the base of the exact results for the lowest cases
$Q=1$ \cite{Samaj1} and $Q=2$ \cite{Samaj2}, let us study
the short-distance stability of the configuration integral
of a neutral cluster formed of one particle of species 1,
placed at the origin ${\vek 0}$, and $Q$ particles of species 2:
\begin{equation} \label{5.4}
I_{\lambda} = \int_{\lambda}^{\Lambda} \prod_{i=1}^Q {\rm d}^2 r_i 
{\prod_{(i<j)=1}^Q \vert {\vek r}_i - {\vek r}_j \vert^{\beta/Q^2} 
\over \prod_{i=1}^Q \vert {\vek r}_i \vert^{\beta/Q} } \ . 
\end{equation}  
Here, $\Lambda$ is a screening length of the Coulomb system
and $\lambda$ is a short-distance cutoff.
After the change of variables ${\vek r}_i = \lambda {\vek r}'_i$,
$I_{\lambda}$ is expressible as follows
\begin{equation} \label{5.5}
I_{\lambda} = \lambda^{2Q - \beta/2 - \beta/(2Q)}
\int_1^{\Lambda/\lambda} \prod_{i=1}^Q {\rm d}^2 r'_i 
{\prod_{(i<j)=1}^Q \vert {\vek r}'_i - {\vek r}'_j \vert^{\beta/Q^2} 
\over \prod_{i=1}^Q \vert {\vek r}'_i \vert^{\beta/Q} } \ . 
\end{equation}  
In the limit $\lambda\to 0$, $I_{\lambda}$ diverges as soon as
$2Q - \beta/2 - \beta/(2Q) < 0$, so that the asymmetric Coulomb 
gas is stable against collapse for $\beta<\beta_{{\rm col}}$, with
\begin{equation} \label{5.6}
\beta_{{\rm col}} = {4 Q^2 \over 1 + Q} \ .
\end{equation}

At the collapse point, although the density of particles (for a given 
fugacity) diverges, the truncated density correlation function 
$\langle {\hat n}({\vek 0}) {\hat n}({\vek r})\rangle^{{\rm T}}$
is finite for $r\ne 0$.
It was conjectured in ref. \cite{Cardy1} and subsequently in
ref. \cite{Jancovici6} that the formula for the second moment of
$\langle {\hat n}({\vek 0}) {\hat n}({\vek r})\rangle^{{\rm T}}$,
having the dimension of $({\rm length})^0$, might be analytically
extended beyond the collapse point.
This second moment remains finite up to the K-T phase transition
(in the limit of vanishing hard-core): its divergence at the K-T point 
is probably associated with a power-law decay of
$\langle {\hat n}({\vek 0}) {\hat n}({\vek r})\rangle^{{\rm T}}$
at asymptotically large distances.
With regard to the sum rule (\ref{4.32}), we therefore suggest that
\begin{equation} \label{5.7}
\beta_{{\rm KT}} = 4 Q \ .
\end{equation}
Note that $\beta_{{\rm col}}<\beta_{{\rm KT}}$, as it should be.
Increasing $Q$, both $\beta_{{\rm col}}$ and $\beta_{{\rm KT}}$
tend to infinity, in full agreement with the fact that there exist
neither the collapse phenomena nor the K-T phase transition for 
the one-component plasma. 

In conclusion, we have studied an asymmetric two-dimensional Coulomb
gas confined to the sphere.
In both canonical and grand-canonical ensembles, the finite-size
expansions of the free energy and of the grand potential are explicitly
shown to possess the same universal term, independent of the
model's details.
The equivalence of both ensembles smears out the previously
indicated difference in the treatment of finite-size corrections
for the one-component and two-component plasmas.
Whether such an explicit calculation of universal finite-size
corrections is possible also for other geometries of confining 
domains is an open problem.

\section*{Acknowledgments}
I thank Bernard Jancovici for valuable discussions and careful
reading of the manuscript.
My stay in LPT Orsay is supported by a NATO fellowship.
A partial support by Grant VEGA 2/7174/20 is acknowledged.

\end{document}